\def\'#1{\if#1i{\accent19\i}\else{\accent19#1}\fi}
\newcommand{\be}{\begin{eqnarray}}
\newcommand{\ee}{\end{eqnarray}}

\documentstyle[preprint,aps]{revtex}
\tightenlines

\begin{document}
\draft
\widetext
\bibliographystyle{unsrt}

\title{ Bloch electrons in electric and magnetic fields}
\author{Alejandro Kunold\footnote{e-mail: akb@hp9000a1.uam.mx}}
\address{Departamento de Ciencias B\'asicas, Universidad Aut\'onoma Metropolitana-Azcapotzalco,  \\
Av. S. Pablo 180, 
 M\'exico D.F. 02200,  M\'exico.}
\author{Manuel Torres\footnote{e-mail: manuel@teorica1.ifisicacu.unam.mx}}
\address{Instituto de F\'{\i}sica, 
   Universidad Nacional Aut\'onoma de M\'exico\\
   Aptdo. Postal 20-364, M\'exico D.F. 01000, M\'exico.}
\date{\today}
\maketitle
\widetext
\begin{abstract}

\hskip0.5cm We investigate   Bloch electrons  in two dimensions subject to constant electric and magnetic fields.   The model that results from our pursuit   is  governed  by   a finite difference  equation with a  quasienergy spectrum  that interpolates between a butterfly-like  structure   and  a Stark ladder structure.   These findings ensued from the  use of electric and magnetic translation operators.

\end{abstract}

\narrowtext

We consider the problem of an electron moving in a two-dimensional lattice in the presence of applied electric and magnetic fields. We refer to this as the  two-dimensional electric-magnetic Bloch  problem (EMB). The corresponding magnetic Bloch system (MB) has a long and rich history. 
An important early contribution to the analysis of the symmetries of the MB problem was made by
Zak  \cite{zak},  who worked out the representation theory of the group of magnetic translations.
 The renowned  Harper equation was derived  assuming a tight-binding approximation  \cite{harp},   and  Rauh derived a dual Harper equation  \cite{rauh} in the strong magnetic field limit.  The  studies of Hofstadter and others  \cite{hof}  of the Harper equation spectrum  have since 
 created   an unceasing  interest in the problem because of the beautiful  self-similar structure of the    butterfly spectrum \cite{review}. A  remarkable  experimental realization of the  Hofstadter butterfly  was recently achieved, not for an electron system, but in  the transmission of microwaves through an array of scatters inside a wave guide \cite{exp}. 
The symmetries of the EMB problem were analyzed some time ago by Ashby and Miller \cite{ash},  who
constructed the group of electric-magnetic translation operators, and worked out their irreducible representations. In this paper we utilize the properties of the  electric-magnetic operators  in order to   derive a finite difference equation that governs the dynamics of the EMB problem. The numerical solution of this equation displays an interesting pattern which interpolates between a butterfly-like structure and a Stark ladder structure.

For the purposes of our research  we consider the  motion of an electron in a two-dimensional 
periodic potential,  subject  to a uniform magnetic field $B$ perpendicular  to the plane and to a constant electric  field $\vec E$,  lying on the plane according to 
$\vec E =  E (\cos \theta, \sin \theta)$ with $\theta$ the angle between  $\vec E$ and the lattice $x_1-$axis. The dynamics of the electron is governed by a time-dependent  Schr\"{o}dinger equation that  for convenience  is  written as

\be
S \,  \Psi   (t,\vec r)   \, = \,    \left[  \pi_0 \,  - \, \frac {\pi_1^2+\pi_2^2}{2 m} \,  - \,   U \left( \vec r \right) \right]
\, \Psi  (t,\vec r) \, =\, 0 
  \label{es}\, ,
\ee
where  $ \pi_0 =  p_0 + A_0$ and $\vec \pi =  \vec p  + \vec A$, 
with the momentum operator  $p_\mu =  \left( i {\partial \over \partial t} ,  - i  \vec \nabla \right)$.  Units have been chosen here in which $\hbar = c  = e = 1$.
Where  necessary we use a covariant notation with space-time three vectors 
$x _\mu = (t,  \vec r); \, \mu =0,1,2$. Eq. (\ref{es}) can be consider as an eigenvalue equation for the operator $S$ with eigenvalue $0$. 
We   adopt a gauge-independent procedure,  thus the gauge potential is written as: 

\be
A_0 &=& \left( \beta - 1\right) \vec r \cdot \vec E \nonumber \, , \\
A_{1}  &=&  \left( \alpha - 1/2 \right)  B   x_2 \,  - \,  \beta E_1   t  \nonumber \, , \\
A_{2}  &=&  \left( \alpha  + 1/2 \right)  B   x_1 \,  - \,  \beta E_2  t 
 \label{gf1}\, .
\ee
This potential yields  the correct  background fields independent  to  
parameters $\alpha$ and $\beta$.   A general potential can be represented by its Fourier decomposition, however  for simplicity we shall consider  the  potential 

\be
U(x_1, x_2) = U_1 \cos\left(2 \pi x_1 /a \right) + U_2 \cos\left(2 \pi x_2  /a \right)
\label{pot} \, .
\ee

 Let $(t, \vec r) \to (t + \tau , \vec r + \vec R)$  be  a uniform translation in space and time, 
where $\tau$ is an arbitrary time and $\vec R $ is a lattice vector.  The   classical 
equations  of  motion remain  invariant under these transformations; whereas  the  Schr\"{o}dinger equation does  not, the reason being the space and time dependence of the gauge potentials.
Nevertheless,    quantum dynamics  of the system  remain invariant under the combined  action of  space-time translations  and  gauge transformations. 
Following Ashby and Miller \cite{ash} we define the electric and magnetic translation operators

\be 
T_0(\tau) = \exp{ (- i \tau {\cal O}_0)}  \, , \qquad
 T_j (a) =   \exp{(i a \,  {\cal O}_j )}
 \label{ope}\, ,
\ee
where  $j = 1,2$ and the symmetry generators are written as new covariant derivatives  
$ {\cal O}_\mu = p_\mu +   {\cal A}_\mu$,  
with   the components of the gauge potentials ${\cal A}_\mu $ given by
\be
{\cal A}_0 &=&  \beta \, \vec r \cdot \vec E \nonumber \,  , \\
{\cal A}_1  &=&  \left( \alpha + 1/2 \right) B x_2 -  \left(\beta -1 \right) E_1 \, t  \nonumber \, , \\
{\cal A}_2  &=&  \left( \alpha  -  1/2 \right) B x_1 -  \left( \beta - 1\right) E_2 \, t 
 \label{gf2}\, .
\ee
The  symmetry operators in Eq. (\ref{ope}) commute with the operator $S$  in Eq. (\ref{es}).  The electric-magnetic operators given by Ashby and Miller  include  simultaneous  space and time translations;  we deemed it  more convenient  to separate the effect of the time evolution  generated by  the    $T_0 $ to that of the space translations generated by   $ T_j$. 
The following commutators can be worked out with  the previous expressions 

\be
 \left[\pi_0,\pi_j \right] = -iE_j  \, , \qquad \qquad
&
&\left[\pi_1,\pi_2 \right] = -i B \, , 
\nonumber \\
\left[ {\cal O}_0,  {\cal O}_j\right] = i E_j \, , \qquad \qquad
&
&\left[ {\cal O}_1,\  {\cal O}_2 \right] = i B \, , 
\label{conmu1}
\ee
with all   other commutators being zero.  
The commutators in  the second line of Eq. (\ref{conmu1})  are part  of the more general  Lie  algebra of the  magnetic-electric 
Euclidean two dimensional group \cite{hadj}.  
 Schr\"{o}dinger's  equation and 
the symmetry operators are expressed in terms of  covariant derivatives
 $\pi_\mu$ and ${\cal O}_\mu$,  respectively. 
A  dual  situation in which the roles  of   $\pi_\mu$ and ${\cal O}_\mu$ are interchanged,  
could  be considered. According to Eqs. (\ref{gf1},\ref{gf2},\ref{conmu1}),  the dual problem corresponds to a simultaneous reverse in the directions  of $B$ and $\vec E$. 

The symmetry operators in Eq. (\ref{ope})  commute with $S$ but  they  do not commute with each other.   We follow a   3  step method to find a  set of simultaneously commuting symmetry operators.  ${\bf (1)}$ First  we consider a  frame rotated  at  angle $\theta$,   with axis along the longitudinal and transverse direction relative to the electric field.  An orthonormal basis for   this frame is  given by
$\hat{e}_L = (cos \theta, sin \theta)$  and  $\hat{e}_T = (-  sin \theta, cos \theta)$.
We assume a   particular orientation of  the electric field, for which  the following condition holds 

\be 
\rho \equiv  \tan{\theta} =  {E_2 \over E_1 } = {m_2  \over m_1  } 
\label{par1} \, ,
\ee
where  $m_1$ and $m_2$   are relatively prime integers.  This condition  insures
that    spatial  periodicity is also found   both along the transverse and the  longitudinal directions. 
Hence, we   define  a rotated lattice spanned by   vectors $\vec b_L = b \hat{e}_L$ and
 $\vec b_T = b \hat{e}_T$ where  $b =  a \sqrt{ m_1^2 +  m_2^2 }$.  The  spatial components of the symmetry generator $\vec {\cal O}$ are projected along the longitudinal and transverse directions :
$ {\cal O}_L =  \hat{e}_L \cdot  \vec{\cal O}$ and
 $ {\cal O}_T =  \hat{e}_T \cdot  \vec{\cal O}$.  It is  readily verified that  $\left[ {\cal O}_0,\  {\cal O}_T \right] = 0$.

${\bf (2)}$   For the rotated lattice,  we regard  the  number of flux quanta per unit cell  to be a rational  number  $p/q$, that is 

\be
\phi \equiv  {B \,  b^2 \over 2\pi } = {p \over q} 
\label{par2} \, . 
\ee
We  can then define  the  extended supperlattice. A rectangle made of $q$ adjacent lattice cells of side $b$ contains an integer number of  flux quanta. 
The   basis vectors of the superlattice   are chosen    as  $q \vec b_L$ and 
$ \vec b_T$. Under these conditions the   longitudinal and transverse magnetic translations  
 $T_L (q b ) = \exp{ (  i q b {\cal O}_L)}$  and   
 $T_T( b ) = \exp{ (  i b {\cal O}_T)}$ 
 define   commuting  symmetries under 
 displacements $q \vec b_L$ and  $ \vec b_T$.  

${\bf (3)}$ We observe that  $T_0$ and  $T_L(q b)$ commute with 
$T_T$.  Yet they fail to  commute with each other: 
$T_0(\tau) \,   T_L (q b )   = T_L (q b)  T_0(\tau) \exp{\left(- i q b \tau   E  \right)}$.
However  the  operators $T_0$ and  $T_L(q b)$  will  commute with one another 
by  restricting    time,   in the   evolution operator,  to  discrete values  with period 
\be  \tau_0 =  {2 \pi \over q b  E} = {1  \over p } \left( {b \over v_d} \right)
\label{period} \, ,
\ee
 where  $v_d =  {E / B}$  and  we utilized Eq. (\ref{par2}) to write the second equality. 
$b/v_d$ is the period of time it takes an electron with  drift velocity $v_d$ to travel between lattice points.
Hence, the meaning of (\ref{period}) is that the ratio of the Stark ladder spacing ($bE$) to the Brillouin zone for the quasienergy $(2 \pi v_d /b)$  is given by the rational number $\phi = p/q$.

We henceforth consider that the three conditions (\ref{par1}),  (\ref{par2}), and  (\ref{period})
hold simultaneously. In this case the three EMB operators: the electric evolution 
${\cal T}_0 \equiv T_0(\tau_0)$ and the magnetic translations ${\cal T}_L \equiv T_L(q b)$, and ${\cal T}_T \equiv T_T(b)$ form a  set of mutually commuting symmetry operators.  
In addition to the symmetry operators, we can define the  energy translation operator  \cite{zak2}

\be
{\cal  T}_E \,    =  \exp{ \left(-i  {2 \pi \over \tau_0} t  \right) }   \, ,  
\label{newop}
\ee
that produces a finite translation in energy by ${2 \pi  /  \tau_0} \equiv q b E$. 
 ${\cal  T}_E $ commutes with the three
 symmetry  operators but not with $S$. Its eigenfunction $\exp{ \left(-i  q b E  \xi  \right) } $ defines a
quasitime $\xi$ modulo  $\tau_0$. 
 
Having defined  ${\cal  T}_0$, ${\cal T}_L$ and  ${\cal T}_T$ that commute with each other and that also commute with $S$,  we can look for solutions of the 
Schr\"{o}dinger   equation  characterized by the  quasienergy ${\cal E}$ and quasimomentum  $k_L$ and $k_T$ quantum numbers according to 

\be
{\cal  T}_0 \,   \Psi  &= & \exp{ \left(-i  \tau {\cal E} \right) } \,  \psi  \, , \nonumber   \\
{\cal T}_L  \,  \Psi  &= & \exp{( i k_L q b )}  \, \psi  \,  , \nonumber \\
{\cal T}_T   \, \Psi  &= & \exp{( i k_T b   )}  \, \psi  
\label{blo1} \, . 
\ee
In particular,  the previous  relations imply  that if  a simple space-time translation acts on  the wave  function, the latter   satisfy the  generalized Bloch conditions 

\be
  \Psi (t + \tau_0, \vec r )  &= & \exp \{-i \tau_0  \left(  {\cal E} +  {\cal A}_0(t,\vec r)  \right)\}   
 \, \Psi (t , \vec r )  
 \,  , \nonumber   \\
 \Psi (t ,  \vec r + q \vec b_L )   &= & \exp\{ i q b \left(  k_L   +  {\cal A}_L(t,\vec r) \right) \}   
\, \Psi (t , \vec r ) \,  , \nonumber \\
 \Psi (t , \vec r  + \vec b_T)   &= & \exp\{ i b \left(  k_T   +  {\cal A}_T(t,\vec r)  \right)\}   
\, \Psi (t , \vec r )  \, .  
\label{blo2}
\ee

We now find convenient to apply a  transformation to new 
variables given by 

\be
X_0&=&- t  \, , \qquad  \qquad \qquad \,\,\,\,\, \,\,\,
P_0 =  {\cal O}_0  \, , 
\nonumber \\
X_1 &=& \pi_T +   m {E \over B}  \, , \qquad \qquad  \,\,   P_1 =  \pi_L/B \, , 
\nonumber \\
X_2&=&  {\cal O}_L/B \, , \qquad \qquad \qquad
 P_2 = {\cal O}_T  + m {E \over B}
\label{can} \, , 
\ee
that satisfy  $\left[X_\mu,P_\mu\right]=i $ for $\mu = 0,1,2$. 
The explicit relation between ($X_\mu \, , \, P_\mu$)  and ($x_\mu \, , \, p_\mu$)  can be worked out using the definition of $\pi_\mu$, and ${\cal O}_\mu$ and Eqs. (\ref{gf1}) and  (\ref{gf2}).
The transformation is not canonical because of   commutator $\left[P_0, X_2\right]=i E/B $,
all   others being zero. 
Applied to   Eq.(\ref{es}) the transformation yields  for the Schr\"{o}dinger   equation 

\be
\left(P_0+ \frac{E}{B} P_2\right)  \Psi    =
\left[  {B^2 P_1^2+X_1^2 \over 2 m } + {m v_d^2 \over 2} + 
U\left(x_1,x_2  \right) 
\right]  \Psi   ,
\label{sch} 
\ee
where   $x_1$ and $x_2$ have to be written  in terms of the new variables
as

\be
{ x_1 \over a}&=& \,\,\,\, {m_2 \over b} \left(P_1  - X_2 - {E \over B} X_0  \right)  \,+\,  
{m_1 \over b B } \left( X_1  - P_2    \right)
\, , \nonumber \\
{ x_2 \over a}  &=& -{m_1 \over b} \left( P_1   -  X_2 -  {E \over B}  X_0  \right) \, +\,  
{m_2 \over b} \left( X_1  - P_2    \right) 
\label{rela}
\, .
\ee
For  $U = 0$  the  dynamics is cyclic in   coordinates $X_2$ and $X_0$.
The appearance of the time variable $X_0$ introduced by the periodic potential suggests
that an approximation  is required   to solve the problem ($e.g.$ adiabatic approximation).
However, as we show below  the use of an appropriate  representation makes such an approximation unnecessary. 
We  adopt the $P_0,P_1,P_2$ representation $\Psi_{ {\cal E}, k_L,k_T}(P) = 
\left\langle P_0,P_1,P_2  \vert  {\cal E}, k_L,k_T  \right\rangle$
with  quasienergy ${\cal E}$ and quasimomentum 
$k_L$ and $k_T$.
In this representation  the $X_\mu$ 
operators act as $X_0 = i \partial / \partial P_0$, $X_1 = i \partial / \partial P_1$ and
$X_2 =  - i  (E/B) \partial / \partial P_0 + i  \partial / \partial P_2$. 
It is readily clear   that the substitution of these relations in Eq. (\ref{rela}) eliminates
 the $\partial / \partial P_0 $ contribution. 
For finding solutions of  Eq.(\ref{sch}) we   split the  phase space $(X_\mu,P_\mu)$ in the  $(X_1,P_1)$ and the $(X_0,P_0;X_2,P_2)$ variables. For the first set of variables 
we choose a set of basis functions in $P_1$ given by the  Landau wave functions
$\chi_n(P_1)$; these functions yield exact eigenvalues of the kinetic part of the 
right hand side of Eq.(\ref{sch}):  $(n + 1/2) \omega_c$,
with the cyclotron frequency $\omega_c = B/m$.
For the subspace generated by the variables $(X_0,P_0;X_2,P_2)$, we notice that the 
four operators (${\cal  T}_0^i$, ${\cal T}_L^j$, ${\cal T}_T^k$,  ${\cal T}_E^l$) with all possible integer values of $(i,j,k,l)$ form a complete set of operators. The demonstration follow similar steps as those  presented by Zak in reference \cite{zaki}.  Hence    a complete set  of functions,  for the subspace $(X_0,P_0;X_2,P_2)$, is provided by the  eigenfunctions of the operators
(${\cal  T}_0^i$, ${\cal T}_L^j$, ${\cal T}_T^k$,  ${\cal T}_E^l$), we write them down and verify their correctness: 

\be
\phi_{k_L,k_T,\epsilon,\xi}(P_0,P_2)   =  \sum_{l,m} c_{m}
e^{i \xi  \sigma  b E \left(p l - m \right) } e^{i (2 \pi /b) m k_L   }  \,\,  
 \delta \left(P_0 -  {\cal E}  -  l q b E   \right) 
 \,\,  \delta\left(P_2 - k_{T} +    \frac{2\pi}{ b } m\right)
\label{wf1} \, , 
\ee
where  we defined  $\sigma \equiv 1/\phi = q/p$. It is easy to check that
this  function automatically satisfies the first and third  EMB tranlations  in Eqs. (\ref{blo1}), whereas the second equation  is satisfied by  imposing   the periodicity condition $c_{m +p} =\,  c_m$.  In addition $\phi$ is also eigenfuction of the energy translation operator
Eq. (\ref{newop}) with eigenvalue $\exp{ \left(-i  q b E  \xi  \right) } $.
The  wave function  $\Psi$ is then expanded in terms of $\chi_n$ and $\phi_{k_L,k_T,\epsilon,\xi}$.
The operator ${\cal T}_E$ is not a symmetry of the problem, so we have to multiply 
$\phi$ by a coefficient $d_\xi$ and add over all possible values of $\xi$; the resulting 
 wave function can be recast as

\be
 \Psi_{k_L,k_T,\epsilon} (P) = \sum_{n,l,m} a_n \chi_n(P_1)\,  h_{m - pl} \,  c_{m} \, 
e^{i \left(2 \pi /b \right)  m k_L  } \,
 \delta \left(P_0 -  {\cal E}  -  l q b E   \right) 
 \,  \delta\left(P_2 - k_{T} +    \frac{2\pi}{ b } m\right)
\label{wf2} \, , 
\ee
where $h_{m - pl} = \sum_{\xi} d_\xi e^{i \xi  \sigma  b E \left(p l - m \right) }$.
 Of particular interest are  the following  Bloch  conditions  obeyed by  $\Psi$  with respect to the eigenvalues:

\be
\Psi\left(  {\cal E},  k_L + b B ,k_T \right) \,  &=& \,  \Psi \left( {\cal E},   k_L,k_T \right)  
 \, , \nonumber \\
\Psi \left( {\cal E},  k_L,k_T + q b B \right)
   &=&  \, e^{i q b k_L}  \, \Psi \left({\cal E}  + q b E,  k_L,k_T \right)
\label{blo3}  \, .
\ee
These conditions are quite different from those satisfied by the usual Bloch  and  magnetic Bloch  functions  \cite{zak};  the second one is  not periodic,  because in addition to   
the Bloch phase   $e^{i q b k_L}$  
the  $k_T \to k_T + qb$ shift   leads to the change  in energy   ${\cal E} \to {\cal E}  + q b E$.

Based on   wave function (\ref{wf2}) it is possible to work out  a complete solution
 of the $EMB$ problem, similar to what has been achieved for the magnetic-Bloch  problem \cite{petschel}. This  will be analyzed elsewhere \cite{nos}. Nevertheless, the approximated solution  wherein the  coupling between different Landau levels is neglected is interesting enough, and probably more illuminating. Within  this approximation the Landau number $n$ is also  a conserved quantum number and   the substitution  of Eq. (\ref{wf2}) in (\ref{sch}) yields the secular equation 

\be 
\left( \Delta -   {E b \over 2 \pi } \Sigma_m  \right)   {\tilde c}_m  =  
f_n(\sigma) \bigg[  U_1  \, \lambda \left(   e^{ - i m_1 \Sigma_m  } {\tilde c}_{m + m_2}   
+  e^{   i m_1 \Sigma_m  } {\tilde c}_{m -  m_2} \right)
 \nonumber \\
 + \, U_2 \,   \lambda^*  \left(   e^{  i m_2 \Sigma_m  } {\tilde c}_{m + m_1}   
+  e^{ - i m_2 \Sigma_m  } {\tilde c}_{m -  m_1} \right)
\bigg] 
\label{he} \, ,
\ee
where  we defined:  ${\tilde c}_m = e^{i \left(2 \pi /b \right)  m k_L  } h_m c_m $, $\Sigma_m = \left( 2 \pi  m -   k_T b \right) \sigma $,  
 $\lambda = \exp \{-i \pi m_1 m_2 \sigma \}$, 
$f_n(\sigma)= e^{- \pi \sigma /2 } L_n\left( \pi \sigma \right)$  and  $L_n$ are 
the Laguerre polynomials.  The quasienergy  ${\cal E}$  is related to 
the eigenvalue  $\Delta$ according  to 
${\cal E}   = \left(n +  1 /2  \right) \omega_c +   m v_d^2 /2 + \Delta$. 

The dynamics of the system is then described by Eq.  (\ref{he}),  a   finite difference equation 
with distant neighbor couplings $m_1$ and $m_2$,  which also includes 
a linear term  proportional  to $\vec  E$.
This equation  generalizes   Harper's equation to include the effect of an   electric field of arbitrary intensity. 
If the    electric field  is switched off, it can be set 
$m_1 = 1$, $m_2 =0$  in which case   Eq.  (\ref{he}) reduces to Harper's equation. 

We present    results for  particular cases  when the electric field is aligned
 along the axis of the original lattice
($i.e.$ $m_1 = 1$, $m_2 = 0$).  It is known that the experimental observation of the  butterfly spectrum  could  possible be  achieved in lateral surface  super lattices \cite{hof,petschel},  
given that    a value for $\sigma \sim 1$  can be obtained for feasible magnetic fields, 
due to  the large dimensions of the   unit cell.  Hence we select  the following  values:   $a \sim 100 \, {\it nm}$, $U_0 \equiv U_1 = U_2 = 0.5 {\it meV} $, $m = 0.07 m_e$ and  $E = 0.05 {\it V / cm}$, that can be satisfied in current experiments  \cite{expe}. 
 For these  values,  $\sigma  = 1$ corresponds to a magnetic field of $10 {\it T}$,
hence  $U_0  / \omega_c \sim 0.05 $ and   the condition required for  weak periodic potential is satisfied.
Figs. (1) and (2) show    plots  for the  scaled   $\Delta$ spectrum as function of $\sigma$.
We recall that for  $\vec E =0$ the spectrum for  $\Delta/ \left[U_0 f_n(\sigma) \right]$  ,   is invariant under the substitution
$\sigma \to \sigma + N$ with $N$ an integer.

\let\picnaturalsize=N
\def\picsize{4.5in}
\def\picfilename{Or10L0.eps}
\ifx\nopictures Y\else{\ifx\epsfloaded Y\else\input epsf \fi
\let\epsfloaded=Y
\centerline{\ifx\picnaturalsize N\epsfxsize \picsize\fi \epsfbox{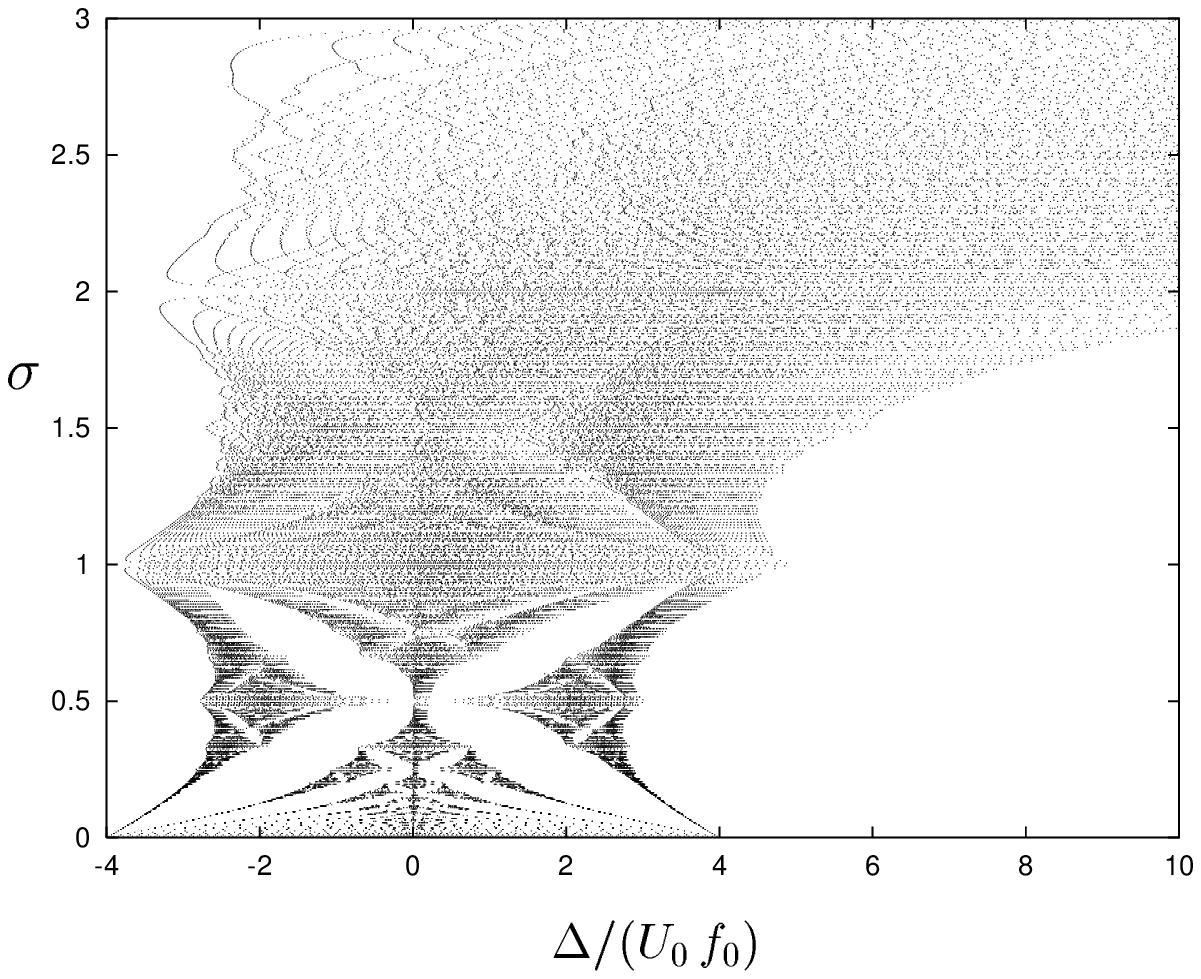}}}\fi

\vskip0.5cm

{\it  Fig. 1  Quasienergy  spectrum  for  the  lowest Landau level.
 The energy  axis is rescaled to $\Delta  / \left( U_0 \, f_0(\sigma) \right)$, 
the parameters  selected are:   $a \sim 100 \, {\it nm}$, $U_0  =  0.5 {\it meV} $, $m = 0.07 m_e$, $E = 0.05 {\it V}/ {\it cm}$ and $k_T = 0$. }

\vskip1.0cm

\let\picnaturalsize=N
\def\picsize{4.5in}
\def\picfilename{Or10L2.eps}
\ifx\nopictures Y\else{\ifx\epsfloaded Y\else\input epsf \fi
\let\epsfloaded=Y
\centerline{\ifx\picnaturalsize N\epsfxsize \picsize\fi \epsfbox{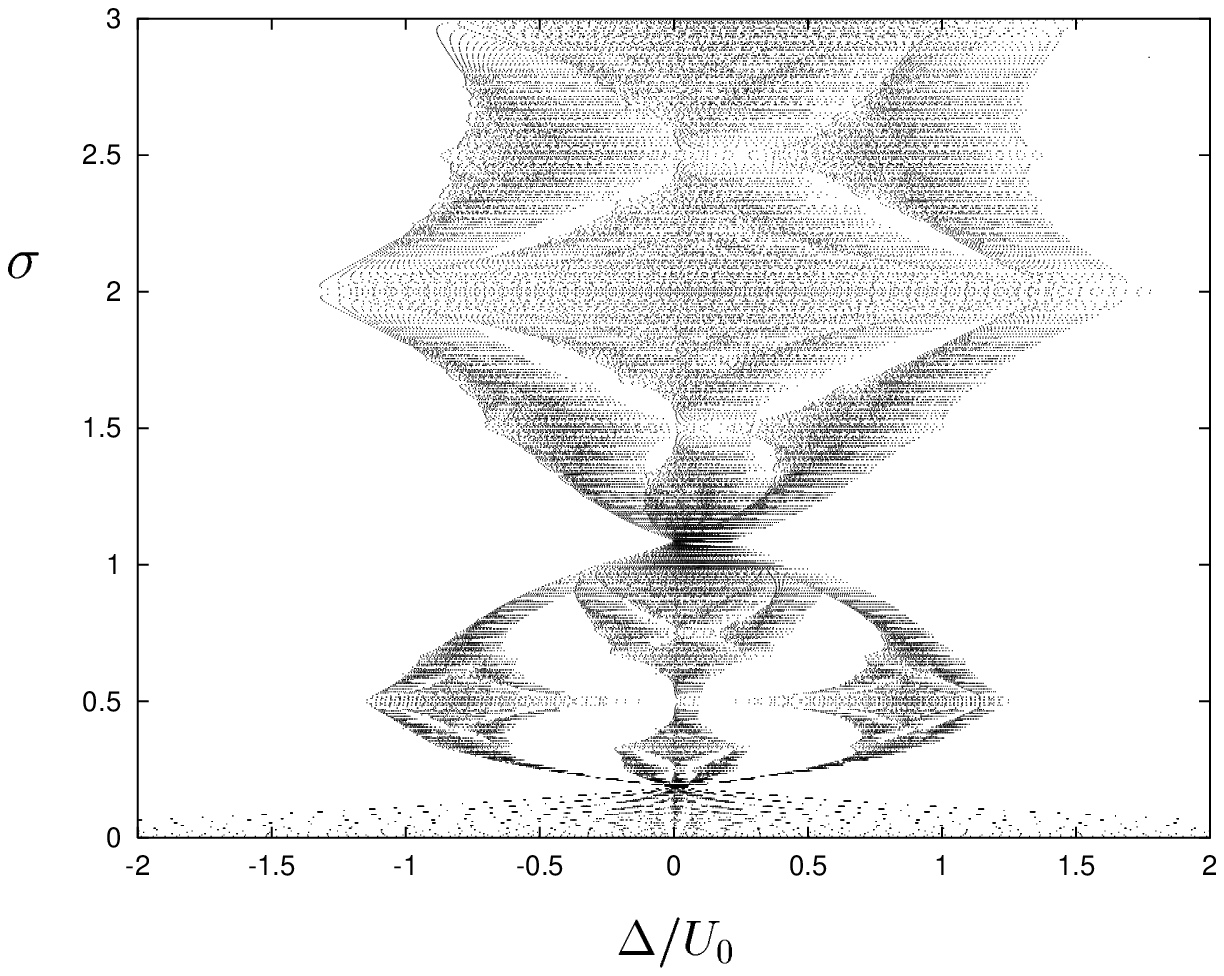}}}\fi

\vskip0.5cm

{\it Fig. 2  Quasienergy  spectrum  of the second Landau level $(n=2)$. In this case the   energy  axis is rescaled to $\Delta  / \left( U_0 \right)$, the other parameters are the same as in Fig. 1.  }

\newpage

In Fig. (1) for the  lowest  Landau level we observe that for a strong  magnetic field
 the Hofstadter butterfly  is clearly shown for $\sigma$ in the interval $[0,1]$.
However, as the  magnetic field  decreases ($\sigma$ increases), some   of the fine-grained structures  of the spectrum are smeared by the effect of the  electric field.
A weak distorted replica of the butterfly is  still observed  for $\sigma$ in the interval $[1,2]$.  
Finally, for bigger values of $\sigma$, 
the spectrum is strongly distorted by  the electric field and  the butterfly ``flies  away''.  At this limit the effect of the  periodic potential is negligible and  the spectrum consists of equally  spaced  levels with separation  
 $  \sigma E b $, $i.e.$ a  Stark ladder.  For  the $n=2$ Landau level  (Fig. 2) 
three smeared replicas of the butterfly can be observed,  the narrow waist  segments  of the plot corresponds to the flat band condition given by 
 $\sigma = \gamma_n / \pi$,  with $ \gamma_n$ a zero of the Laguerre polynomial.

In conclusion,  the electric and magnetic translation symmetries are utilized to analyse  the EMB problem.  Bloch functions are derived and their properties established in Eqs. (\ref{blo2}) and 
(\ref{blo3}). The system is  governed  by    Eq. (\ref{he}),   the   spectrum  of which  interpolates between a butterfly-like  structure   and  a Stark ladder structure.  
This equation  offers a very interesting model,  susceptible of analysis in terms of    
dynamical systems.   We finally remark  that    the present formalism  
should set    the basis  for the study 
of  Hall conductivity beyond the linear response approximation.

\acknowledgments  
We have profited from helpful  discussions with Roc\'io J\'auregui and  Jos\'e Luis Mateos.

\end{document}